# Strongly-Correlated Electron-Photon Systems

Jacqueline Bloch, Andrea Cavalleri, Victor Galitski,[*] Mohammad Hafezi, and Angel Rubio

**An important goal of modern condensed matter physics involves the search for states of matter with new emergent properties and desirable functionalities. Although the tools for material design remain relatively limited, notable advances have been recently achieved by controlling interactions at hetero-interfaces[1], precise alignment of low-dimensional materials[2] and the use of extreme pressures[3]. Here, we highlight a new paradigm, based on controlling light-matter interactions, which provides a new way to manipulate and synthesize strongly correlated quantum matter. We consider the case in which both electron-electron and electron-photon interactions are strong and give rise to a variety of novel phenomena. Photon-mediated superconductivity, cavity-fractional quantum Hall physics and optically driven topological phenomena in low dimensions are amongst the frontiers discussed in this perspective, which puts a spotlight on a new field that we term here "*strongly-correlated electron-photon* science."**

## 1. Background

A remarkable convergence is taking place at the interface between traditionally separate areas: quantum materials, quantum optics, and nonlinear laser physics. Important progress has already been made in controlling the coupling between single photons and quasiparticle excitations in solid-state systems[4,5]. These advances include the optical control of many collective modes of solids, including excitons, plasmons, phonons, skyrmions, magnons, solitons, and individual electrons. These efforts are already impacting research in new materials and devices for future quantum technologies, such as quantum memories, transducers, and networks, and materials for energy and sensing applications. However, this research has almost exclusively relied on materials where electron correlations are weak.

At the same time, strong optical fields have been employed to drive, switch, or create new cooperative responses in materials with strongly interacting electrons. Initial demonstrations have ranged from optically enhanced high-temperature superconductivity[6] to photo-induced magnetism[7], ferroelectricity[8] and non-equilibrium topological quantum phases[9,10]. In all these experiments, the strong photo-susceptibility of correlated electron systems is exposed, but the electromagnetic field remains classical, hence covering only a fraction of what is possible. Yet, these optical control experiments provide a preliminary sample of a variety of physical phenomena to come when the quantum mechanical properties of light are brought to the fore.

Another recent research strand has involved an extension of the cavity-QED toolbox to complex material systems, with the goal of engineering hybrid light-matter phases[11,12,13,14,15,16]. New predictions have already been made for light-matter hybrids based on quantum materials, where electron and photon interactions contribute to the emergent physical phenomena on an equal footing. Beyond initial experimental demonstrations[17] and promising theoretical proposals[18,19,20], we highlight the enhancement of superconductivity[21,22] and ferroelectricity[23] in cavities as an especially attractive possibility.

---

[*] Corresponding author

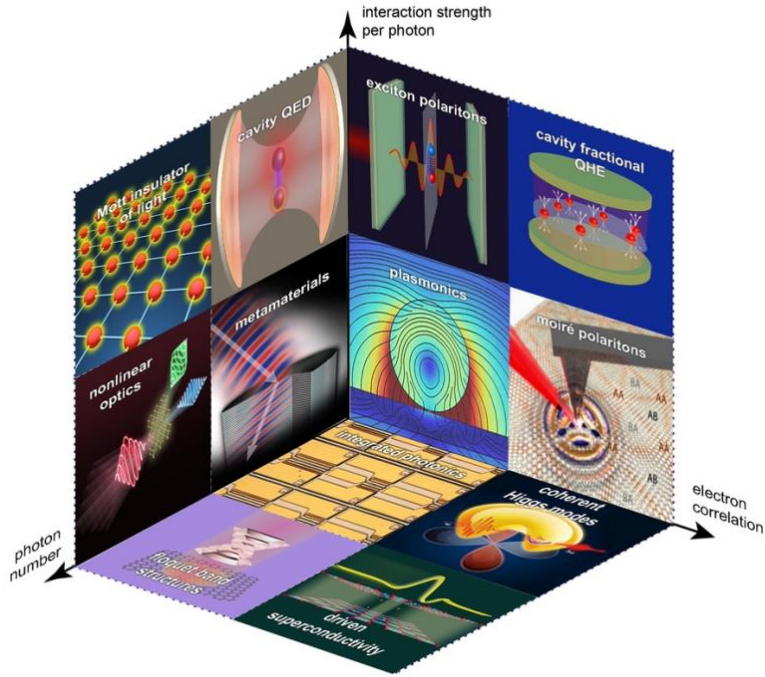

*Figure 1 **Regimes in quantum optics. Top:** Map for strongly correlated electron-photon systems. We classify the many emerging quantum phenomena in terms of three physical parameters: the strength of the light-matter coupling, the strength of electronic correlations and the photon number (light intensity). **Left vertical panel:** Phenomena expected in systems in which electronic correlations are weak. For weak light-matter interaction lower left and region around the corner, linear optical phenomena in metamaterials evolve into nonlinear optical responses as the photon number is increased. On the same panel, as light-matter coupling becomes strong (vertical axis) one finds the strongly correlated photonic phenomena in cavity QED. For strong coupling and large photon number, these evolve into Mott insulators at intense light fields. **The vertical panel on the right** spans the strength of electronic correlations. We find nonlinear spectroscopies such as time-domain Higgs-mode detection in superconductors as well as cavity fractional Quantum Hall physics. **The floor panel**: includes include linear (ARPES) and nonlinear spectroscopy of materials (SHG polarimetry) at various degrees of electronic correlations, as well as driven materials with weakly correlated electrons, in which Floquet band structures emerge. Finally, for strong fields and strong electronic correlations, photo-induced emergent phenomena in superconductors, magnets, and ferroelectrics are expected.*

Figure 1 provides a conceptual three-dimensional map of the variety of phenomena possible at the intersection of material research, quantum optics, and laser physics. The three axes represent the strength of the light-matter interaction (per photon), the strength of electronic correlations, and the intensity of light. An incomplete selection of examples is intended to illustrate the breadth of phenomena included in this three-dimensional space. Starting at the origin of the plot, where all parameters are small, we find the physics of metamaterials, plasmonics, photonic circuits. By increasing the photon number, whilst keeping both electron correlations and light-matter coupling weak, one finds nonlinear optical phenomena[24] and Floquet engineered band structures in semiconductors and semimetals[25]. Along the vertical axis, as the interaction strength per photon is increased, one enters the cavity-QED regime[26], where one photon can block transmission of another[27]. By increasing the number of photons, the so-called quantum fluids of light can be obtained, for example in the form of exciton-polariton condensates. For large light-matter coupling *and* many photons, strongly correlated photonic states such as a Mott insulator of light can be realized[28].

Following the electron correlation axis leads us into a regime where quantum materials are probed by nonlinear optical techniques. For example, the dynamics of the order parameter can be studied in time-domain spectroscopy, as is the case for coherent Higgs modes in superconductors[29] or Moire excitons in twisted van der Waals materials[30].

This perspective focuses on a subset of these regimes. As a background, we start by discussing weakly correlated materials that are periodically driven by strong electromagnetic fields. These are discussed in the context of Floquet band-structure engineering (section 2). We then focus on several applications in the control of quantum materials, which range from superconductivity to ferroelectricity and magnetism (section 3). Finally, we discuss situations where both large light-matter coupling and strong electron-electron correlations are present, where interesting cavity-enhanced or cavity-induced collective phases could appear. Examples include possible creation of cavity-mediated superconductivity and polaronic fractional quantum Hall states (section 4).

**2. Floquet Band Structures in Electronic Materials**
Periodic-in-time driving provides a qualitatively new type of quantum control of materials, especially in regimes in which the interaction of light with a solid is both *non-linear* and *non-perturbative*. This is a situation in which the light-matter coupling acts as a leading term in the system's Hamiltonian and is to be considered on an equal footing with electron kinetic energy, electron interactions, and crystal fields. Periodically driven systems are often described by Floquet theory, dating back to 19th century work by Gaston Floquet on differential equations with periodic functions. The Floquet formalism provides a straightforward description of how a periodic optical field renormalizes the spectrum of excitations of a solid. In the same way that a spatially periodic crystal structure creates replicas of free electron dispersion, a time-periodic modulation creates replicas of the energy spectrum[31,32]. The emergence of a renormalized band structure, which in the limit of high frequency modulation is conveniently cast in terms of a time-independent Floquet Hamiltonian, has been demonstrated to emerge in a variety of contexts from synthetic optical lattices[33] to time crystals[34].

For most experimentally accessible protocols, the relation amongst the underlying material band-structure, the light-matter coupling, the optical drive and the effective Floquet Hamiltonian is not simple. Nevertheless, it has been shown theoretically that even a non-remarkable constituent Hamiltonian can lead to remarkable (e.g., topological) Floquet band structures when subjected to a periodic drive (Figure 2a). Recent experimental advances have validated these ideas in real materials. The pioneering work by Gedik and co-workers has demonstrated the existence of Floquet-Bloch states in the topological insulator $Bi_2Se_3$ (Figure 2b) where a circularly polarized light pulse, tuned to a frequency below the bulk gap to minimize dissipation, gave rise to Floquet-Bloch bands. Another notable advance has been the development of experimental tools to measure ultrafast transport in Floquet-driven states, as demonstrated in a recent experiment on a light-induced anomalous Hall effect in monolayer graphene (see Figure 2b).

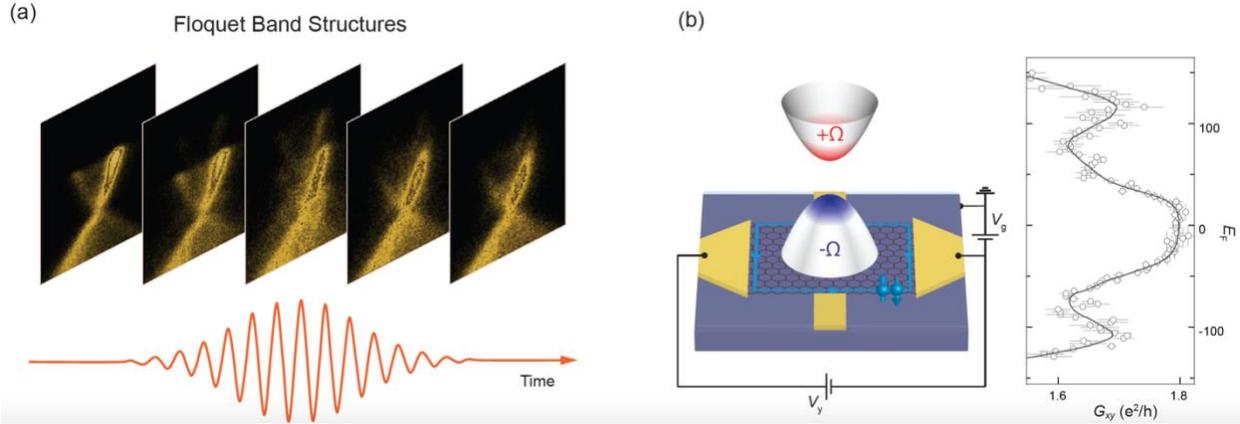

*Figure 2 **Floquet topological physics.** (a) Observation of the Floquet-Bloch states in time-resolved ARPES experiments in the topological insulator $Bi_2Se_3$ (ref. Wang et al.) (b) Observation a light-induced anomalous Hall effect in a monolayer graphene, measured by femtosecond electronic techniques (ref. Mciver et al.)*

Understanding the role of dissipation remains a challenge, as even in the simplest situations[35,36,37,38] decoherence and heating of the material play a fundamental role in the population and stabilization of Floquet states. A particularly intriguing research direction involves Floquet states with no equilibrium counterpart, which until recently have only been discussed primarily in the theoretical literature. Yet, experiments are beginning to realize these anomalous Floquet states, either in the synthetic environment of ultracold gases and photonic simulators, or in real materials, such as in two-dimensional homo and hetero-structures. For example, the chiral Floquet phase[39] has been proposed to be a driven quantum-Hall-like state. The proposed protocol involves the renormalization of two bands connected by a chiral edge state. When these bands move around the Floquet energy circle, they collapse onto one another and create a trivial Floquet band structure in the bulk. However, the edge state cannot be eliminated and yield an example of a non-equilibrium topological state without an equilibrium analogue. This anomalous Floquet state was recently realized with cold atoms and coupled photonic waveguides[40], but its realization in solids has not happened yet. Other remarkable proposals have been made, including "Floquet quantum criticality"[41] via many-body localization in disordered systems.

One current frontier is to define, classify, and potentially realize interacting Floquet phases. Some theoretical progress has been made in understanding driven symmetry-protected topological (SPT) phases. There exists a sophisticated mathematical formalism[42] to classify equilibrium interacting quantum phases protected by a symmetry group, $G$. Periodic driving effectively enriches the symmetry group by the discrete time-translational symmetry, $Z$, enlarging it into $G \times Z$. This changes the classification scheme and gives rise to new phenomena[43]. Another related phenomenon can be realized in certain driven spin chains[44] in which boundary spins flip every period and show period doubling. This behavior, dubbed a Floquet time crystal[45], represents the spontaneous breaking of discrete time-translation symmetry. It was observed shortly after its prediction with ions[46] and also in a diamond crystal doped with NV centers[47].

Another interesting direction is the possibility of using Floquet engineering to control the strength and form of interparticle interactions. Examples include engineering interaction in Mott insulators[48,49], three-body interactions in synthetic systems and in the context of fractional quantum

Hall effect (FQHE). Specifically, in the case of Floquet FQHE, it has been theoretically proposed that by resonantly selecting and driving the optical transition between Landau levels, one can form a synthetic bilayer quantum Hall state in a monolayer system. In this case, the relevant Coulomb interaction terms take a non-trivial form of great interest to quantum Hall physics[50] and may help stabilize exotic topological states, such as the highly sought-after Fibonacci phase[51]. If one can modulate the strength of two-body interactions, three-body and higher-order interaction terms could also be generated[52]. Many questions remain open such as the possibility of topologically-ordered Floquet phases, generalizing fractional quantum Hall states and toric code-like models to driven systems, as well as the possibility of topological defects in Floquet dynamical systems and gauging the time-translation symmetry.

## 3. Driven Strongly Correlated Electrons in Functional Materials

The response of strongly correlated materials to optical irradiation has been studied for decades, although the appearance of TeraHertz and mid-infrared sources of sufficient power has drastically enriched this field. Indeed, strong light fields at these frequencies have been used to drive nonlinearly the low-energy degrees of freedom that determine key emergent properties of solids already in equilibrium (Fig. 3). Nonlinear excitation of collective modes has provided a new set of protocols, sometimes creating states with no equilibrium counterpart. For instance, a set of applications of strong-field driving of quantum materials dealt with the response of cuprate superconductors to single-cycle TeraHertz pulses, demonstrating superconductor-metal oscillations[53], parametric amplification of the superconductivity[54], and revealing superfluid charge stripes[55]. Similarly, high electric field transients have been used to drive an insulator-metal transition in a $VO_2$ metamaterial structure[56], to manipulate magnetic dynamics[57], soft modes in incipient ferroelectrics, and topology in Weyl semimetals[58].
Another strand of research has involved the study of directly driven crystal lattices, where certain phonons are excited nonlinearly[59] to deform and dynamically modulate the equilibrium structure of a solid and with it electronic[60], orbital[61] and magnetic orders. Notable were experiments where the atoms in the solid were driven in loops, creating a Floquet phase with broken time-reversal symmetry[62], now supplemented by recent experiments in which a ferrimagnetic polarization was achieved via control of the crystal field [63].

Amongst all of these cases, arguably the case of periodically driven *superconductors* is the one that has raised most interest and posed some of the most provocative questions. Conventionally accepted wisdom posits that external non-equilibrium perturbations destroy quantum-coherent states of matter via heating. This is often so, but not always. A striking example, dating back to the 1960s, involves the enhancement of superconductivity in microbridges exposed to microwave irradiation[64,65,66]. The response of these superconductors exhibited higher transition temperatures and critical currents than in equilibrium. This effect has been interpreted as a result of quasiparticle redistribution[67,68], based on the observation that the key equation in the theory of superconductivity depends on the distribution function of electrons. As external radiation reduces the quasiparticle population near the superconducting gap, non-equilibrium superconductivity is favored (Fig. 4a).

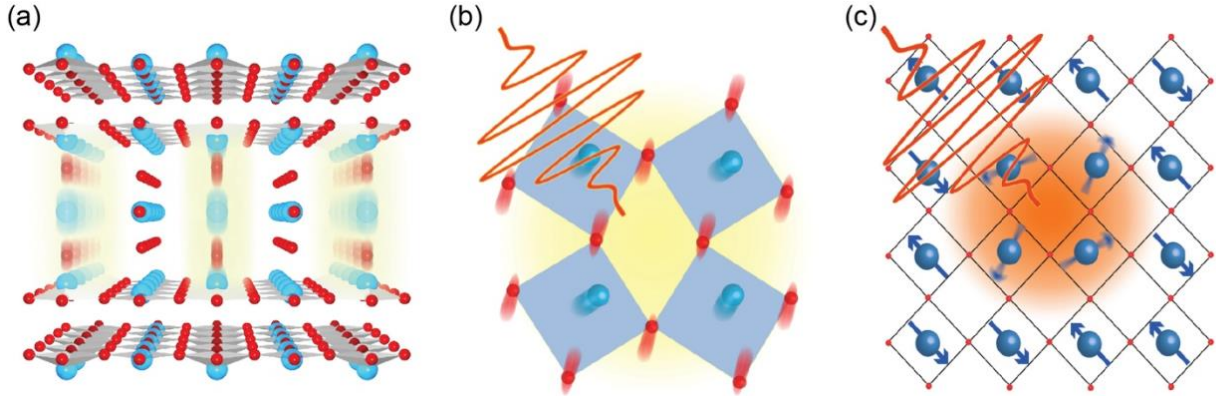

*Figure 3* **Driven quantum material**. *(a) superconductors (b) ferroelectrics (c) ferromagnets* have been manipulated by applying a time-dependent optical or TeraHertz field, which has been used to redistribute electrons amongst bands, distort the lattice through nonlinear phononics or by introducing a time-varying potential that dynamically renormalizes the energy landscape of the system.

Many of these ideas are now coming back to the fore in new contexts, especially in unconventional superconductors driven using near-infrared[69,70] and mid-infrared frequency fields, instead of microwaves. Nonlinear phonon excitations have been applied to cuprate materials at doping levels where superconductivity competes with alternative forms of charge and spin order, such as in striped $La_{1.675}Eu_{0.2}Sr_{0.125}CuO_4$, but also in the bilayer high-$T_c$ cuprate $YBa_2Cu_3O_x$ upon modulation of the apical-oxygen atom positions. Most strikingly, the temperature scales up to which superconducting-like optical spectra were observed followed the mysterious pseudogap transition line T*, extending above room temperature in most underdoped compounds.

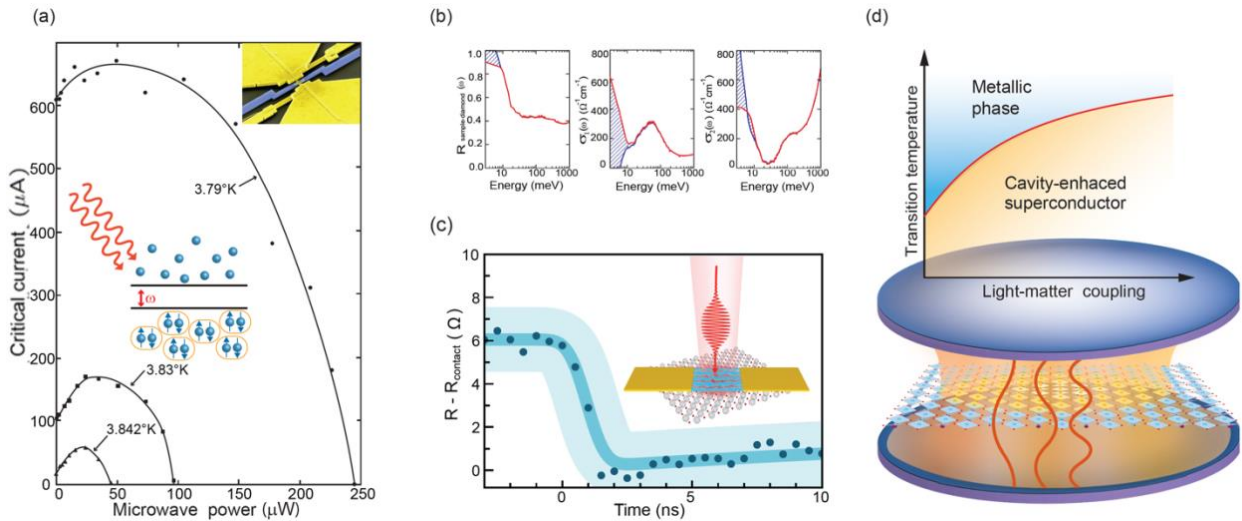

*Figure 4.* **Optical enhancement of superconductivity**. **4a** Enhanced critical current as a function of irradiating power in microbridges of BCS superconductors. Irradiation enhances the properties of the equilibrium superconductor by promoting quasiparticles from the gap region to higher energy states. **4b.** Transient optical spectra recorded in $K_3C_{60}$ upon irradiation with femtosecond pulses at 6.7 μm wavelength. These optical properties are observed up to temperatures of at least 150 K, closely following those of the equilibrium superconducting state observed in the same material only for $T<T_c$=20 K (Ref. Mitrano et al.). **4c.** Transient electrical response of $K_3C_{60}$ *irradiated* with nanosecond pulses at 10.6 μm wavelength. A zero-resistance state, which survives for at least 10 ns is observed in these experiments (Ref Budden et al.).

Similar phenomena have now been reported for materials other than cuprates, such as the $K_3C_{60}$ superconductor[71] (Fig. 4b). This state was recently shown to be metastable when driven at sufficient amplitudes and for a sufficiently long time, opening new opportunities to generate photo-induced high-temperature superconductivity in steady state[72]. In a related experiment, driven K-BEDT-Br[73] was also shown to exhibit superconducting-like properties, which however appeared to be different from those of the equilibrium superconductor observed when cooling below $T_c$.
The first attempt to describe these phenomena theoretically has followed the idea of light-induced melting of a competing charge order, which is expected to enhance superconductivity[74]. These early papers were followed by other studies that proposed the possible role of amplification of the superconducting instability[75], the possible cooling of phase fluctuations[76], laser-induced electron attraction[77], and a plasmonic mechanism of melting the competing order[78]. Despite this intense work, the theoretical jury is still out on the underlying microscopical nature of observed light-induced superconductivity.

## 4. Strong Light-Matter Coupling in Cavity-QED

Light-matter interaction can be enhanced by the confinement of electromagnetic modes within small-volume resonators, and by increasing the quality factor such resonators. We label this situation in Fig. 1 with the "interaction strength per photon" axis. Such enhancement can lead to qualitatively new physics, revealing phenomena that have no free-space counterparts, regardless of the strength of the laser field.

Historically, until the late 1970's, non-linear optical effects were confined to the classical regime and were only accessible at high optical powers. In order to increase the nonlinear effects and make them important at the level of individual photons, the field of "cavity-QED" was invented. The basic idea is to place matter, initially made up of individual atoms, between two metallic mirrors that form a cavity. The nonlinear response in such a system can be so large that it can dramatically change the statistics of photons going through the cavity and even influence the transmission of a single "target" photon with a "control" photon. The development of cavity-QED was fueled by growing interest in quantum information sciences to achieve photonic quantum logic. The pioneering works in the optical domain used trapped alkaline atoms, and then solid-state emitters, such as quantum dots and color centers in diamond. In the microwave domain, the pioneering works have involved Rydberg atoms, and later, "artificial atoms" using Josephson junctions (for a review on these topics see Ref 79 and 80).

Until recently, most of the focus in cavity-QED systems remained on the single spatial modes of photons. For example, although proposed in early theoretical works[81], strong quantum effects in propagation, such as the formation of two-photon bound states, were observed only recently[82]. One expects that the combination of cavity-QED and the correlated electronic states in materials can lead to a new paradigm for creating, probing, and manipulating strongly correlated states and novel light-matter hybrids[83,84,85]. Specifically, this is expected in the strong coupling regime, where electronic excitations such as excitons, strongly couple to light and form polaritons. The resulting effect of light-matter interaction is no longer perturbative, and it dramatically modifies the physical properties of the system. We review recent developments and discuss future opportunities.

One example of strong light-matter interaction is based on exciton-polaritons[86] that can be viewed as quasi-particles resulting from the strong coupling between photons confined within an optical cavity, and excitons in a quantum well (Fig. 5a). The hybrid light-matter quasi-particle exhibits a wealth of interesting physical properties inherited from its mixed nature. Because of the coupling of photons trapped in the cavity with the electromagnetic modes of the outside world, the system is intrinsically dissipative, and engineering of the pump field is a key knob to manipulate excitations. Importantly, excitonic polaritons exhibit a giant third-order (Kerr) non-linearity, which originates from Coulombic exchange interaction between the fermionic constituents (electrons and holes) of excitons. As a result, polaritons behave as interacting photons exhibiting many fascinating phenomena, to some extend similar to cold-atom systems, such as Bose-Einstein condensation[87,88], superfluidity[89] with subtle physical properties related to their driven dissipative nature[90,91,92], or nucleation of topological defects). Moreover, an enhancement of conductivity in organic semiconductor has recently been observed in strong coupling regime with plasmonic modes[93]. Interestingly, such quantum fluids of light provide a testbed for the exploration of out-of-equilibrium condensates in a driven-dissipative context. Most of the polariton features studied so far have been successfully described within a semi-classical mean-field approach, neglecting quantum fluctuations and correlations. However, the true quantum regime is now within reach. Very recently, weak anti-bunching of light transmitted through a polariton cavity was reported[94,95]. Various proposals are currently being pursued to enhance polariton interactions deeper into the quantum regime. The main idea is to couple light to different kinds of excitations that have stronger interactions. Examples include dipolar excitons[96], polaron-polaritons[97] and fractional quantum Hall states[98].

A further research direction in this area involves engineering photonic lattices. Confined modes of individual resonators hybridize to form bands of extended states as atomic orbitals do in crystals. Depending on the symmetry of the lattice, photon properties can be fundamentally modified and emulate interesting Hamiltonians. Recently, pioneering experiments have demonstrated the feasibility of these ideas in a variety of photonic systems such as coupled waveguides, ring resonators, photonic crystal cavities[99,100,101,102] Meanwhile, epitaxially grown polaritonic materials are particularly interesting since in addition to engineering complex potential landscapes in a driven-dissipative non-linear system[103,104], the resulting collective modes can be directly accessed both in real and reciprocal space. Such synthetic photonic materials have allowed simulating various condensed matter models. A particularly exciting new development in this research is the emergence of the field of topological photonics[97]. More generally, these efforts have opened a wide playground with no electronic counterpart. Already in the weakly nonlinear regime, examples include strong artificial gauge fields that cannot be naturally accessed, very large strains can be emulated, and Floquet Hamiltonians than can be simulated by modulation in the propagation direction. These ideas have already led to proof-of-concept demonstration in using topology as a design principle for robust optical devices, such as topological lasers[105,106,107] and topological quantum sources of light[108].

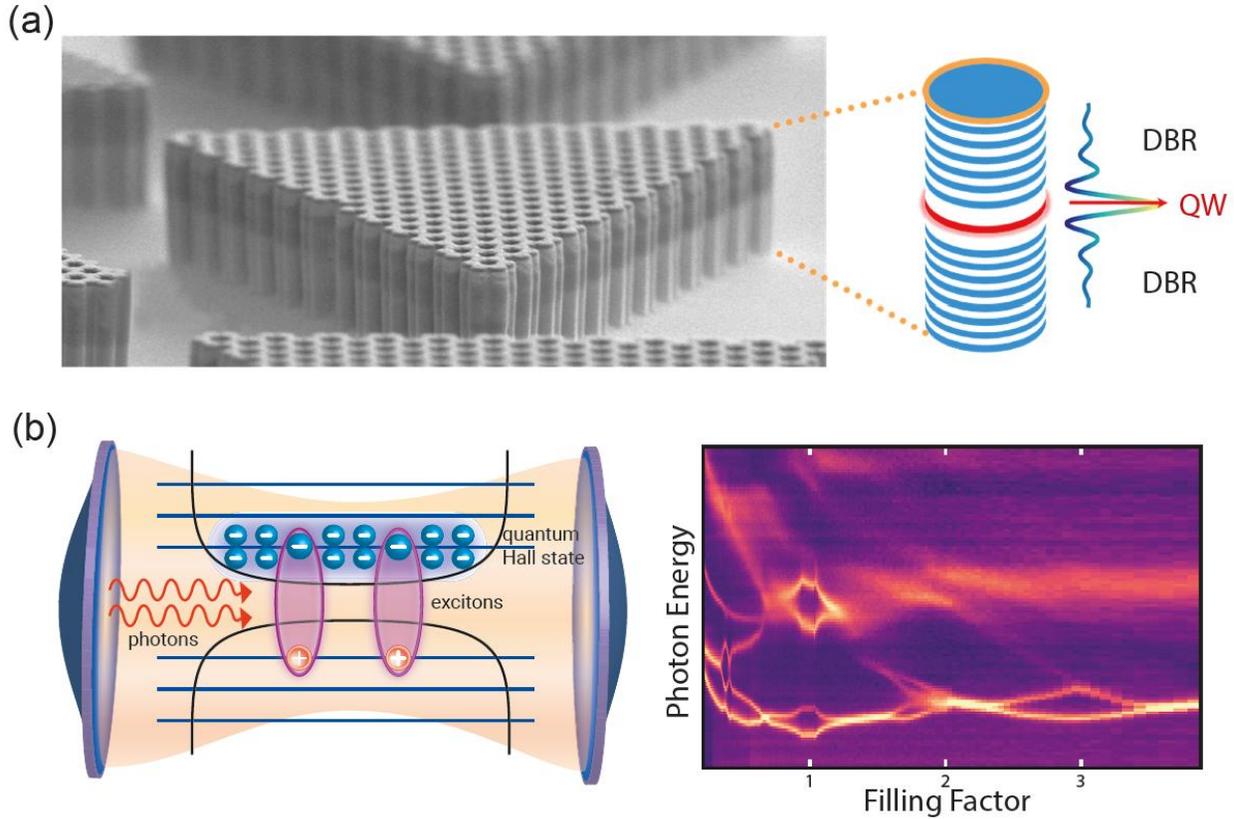

*Figure 5 **Correlated material in optical cavities**. (a) Scanning electron microscopy image of honeycomb lattices of polariton cavities; the sketch on the right depicts the layers within each cavity made of two distributed Bragg reflectors (DBR) surrounding a cavity with active material (such as a quantum well (QW)) at its center (Courtesy to A. Amo and J. Bloch)(b) schematic representation of an electrically gated cavity structure containing a two-dimensional electron gas operated in a quantum Hall regime. Optical resonances measured in reflectivity as a function of the 2DEG filling factor which is varied with the magnetic field (Courtesy to A. Imamoglu, P. Knüppel and S. Ravets) (c) A schematic of a theoretical proposal for cavity-enhanced superconductivity. Embedding a superconductor in a cavity, where the vacuum fluctuations are contained in highly reduced mode volumes, is expected to mediate pairing through cavity photons.*

Another interesting route is to couple complex photonic lattices with a strongly interacting material. One could envision a new research field where a scalable ensemble of coupled cavities in the blockade regime could be created. Then, strong correlations could be imprinted on the active material via the coupling to the photon field. Excitation schemes to generate spatially entangled multiphoton states[109], or to stabilize multi-photon Laughlin states would be of high relevance[110]. Multiphoton pumping schemes[111], multiphoton loss mechanisms and more generally reservoir engineering[112] are among the possible tools to create and stabilize strongly correlated many-body quantum states. Beyond fundamental interest, this research may find applications in quantum information technologies, such quantum sources of light, photonic quantum logic, and analogue quantum simulators.

A particularly attractive direction here is enhancing superconductivity in an optical cavity (Fig.5c). For example, if electron-photon coupling can be renormalized by the cavity, one may reach long-lived stimulated superconductivity at increased temperatures. Furthermore, it has been proposed that transverse fluctuations of the electromagnetic field could mediate pairing, a phenomenon that has vanishing amplitude in free space but one that can be enhanced once the coupling to cavity fluctuations is made very strong. Finally, it has been proposed that by coupling a cavity to a superconductor, one can redistribute quasiparticles into a more favorable nonequilibrium distribution and promote pairing and therefore enhance the superconductivity[12]. Experimentally, these effects are largely unproven although an anomalous diamagnetic response was reported in the normal state of $Rb_3C_{60}$ when this material was embedded in a cavity, potentially related to incipient cavity-mediated superconductivity. Another interesting direction is the exploration of a cavity-induced ferroelectric phase transition[23,113] which aims to realize the paraelectric-ferroelectric transition in $SrTiO_3$, already achieved by strain, isotope substitution and light, to be realized in a cavity by coupling the order parameter to the cavity modes.

Another intriguing direction is to investigate the cavity-QED physics of other correlated light-matter states, such as fractional quantum Hall states, and explore the resulting quantum optical response of the system (Fig. 5b). Specifically, it has been experimentally observed that at low carrier density, the polariton-polariton interaction strength depends on the nature of underlying many-body electronic state[98,00], while the numerical simulation suggests that for high-density of carriers such dependence could be washed out[114]. Therefore, the effect of strongly correlated electronic states on the quantum optical response and engineering strong effective photon-photon interactions remain open problems. A challenge would be to optically excite, probe and manipulate anyons, which are fractionally charged quasiparticles in quantum Hall systems[115]. Such ideas could be applied to both two-dimensional electron gases and van der Waals materials[116]. A promising class of strongly coupled light-matter systems is atomically thin monolayers and bilayers such as twisted bilayer graphene and transition metal dichalcogenides. These systems have recently emerged as fundamental platforms to enhance correlations[18,117]. Combining twist-, non-equilibrium-control and cavity-QED offers opportunities to realize novel correlated states of matter touching most of the regimes in Fig.1.

Moreover, strong light-matter coupling in a cavity provides a way to create states with broken symmetries, such as time-reversal symmetry achieved by nonlinear optical response[118] or coupling to quantum fluctuations in chiral cavities[14] In this context, there have been theoretical proposals where the cavity dressing results in the desired changes to the ground state of materials and could realize non-equilibrium states of matter that have been seen in laser-driven materials. Whilst the physical processes behind laser and cavity-dressing of matter are different, the properties of such dressed-hybrid states can be similar, since they descend from formally similar Hamiltonians.

## 5. Outlook
In this Perspective, we addressed a selected number of new directions in exploiting light-matter interaction to control and create emergent phenomena in solids. On the one hand, this new field draws inspiration from conventional design efforts in condensed matter research, with this analogy being most evident in the use of TeraHertz pulses to dynamically deform crystal lattices and engineer the atomic structure. On the other hand, the rich physics of cavity-QED, quantum fluids of light, and quantum optics in general, when combined with strongly correlated electron materials provide an entirely new view of light-matter coupling. Progress in all these fields requires the development of new theoretical frameworks to describe strongly interacting light-matter phenomena. Learning from the progress in the ultra-cold atom platform, where mastering light-matter interaction at the level of a single quantum led to the development of synthetic quantum matter, one expects a similar revolution in condensed matter systems coupled to light or cavities. Given the recent progress in both theory and experiment to harness light-matter interaction with ever-increasing precision, a march towards strongly correlated electron-photon matter is foreseeable.

**Acknowledgements**

We gratefully thank J. Curtis for critical reading of the manuscript. V.G. was supported by NSF DMR-2037158, US-ARO Contract No. W911NF1310172, and Simons Foundation. MH acknowledges support from AFOSR FA95502010223, FA9550-19-1-0399, ARO W911NF2010232, and ARL W911NF1920181 and Simons Foundations. JB acknowledges financial support from Paris Ile-de-France Région DIM SIRTEQ, H2020-FETFLAG project PhoQus (820392), QUANTERA project Interpol (ANR-QUAN-0003-05), ANR project Quantum Fluids of Light (ANR-16-CE30-0021), and the French RENATECH network. AR is supported by the European Research Council (ERC-2015-AdG-694097), Grupos Consolidados (IT1249-19) and the Flatiron Institute, a division of the Simons Foundation. We acknowledge funding by the Deutsche Forschungsgemeinschaft (DFG) under Germany's Excellence Strategy - Cluster of Excellence Advanced Imaging of Matter (AIM) EXC 2056 – 390715994 and by the Deutsche Forschungsgemeinschaft (DFG, German Research Foundation) -SFB-925 - project 170620586. Support by the Max Planck Institute – New York City Center for Non-Equilibrium Quantum Phenomena.

**Authors contribution**
All authors discussed the material and wrote and revised the manuscript together.

**Authors Information**
Jacqueline Bloch, Centre de Nanosciences et de Nanotechnologies (C2N), Universite Paris Saclay - CNRS, 911200 Palaiseau, France, jacqueline.bloch@c2n.upsaclay.fr

Andrea Cavalleri, Max Planck Institute for the Structure and Dynamics of Matter, Hamburg, Germany, andrea.cavalleri@mpsd.mpg.de

Victor Galitski, Joint Quantum Institute and Department of Physics, University of Maryland, USA
galitski@umd.edu

Mohammad Hafezi, Departments of Physics and ECE, University of Maryland, USA
hafezi@umd.edu

Angel Rubio, Max Planck Institute for the Structure and Dynamics of Matter, Hamburg, Germany
Center for Computational Quantum Physics (CCQ), Flatiron Institute, New York NY 10010
arubio@flatironinstitute.org